\documentclass[letterpaper]{article} 
\usepackage{aaai22}  
\usepackage{color}
\usepackage{times}  
\usepackage{helvet}  
\usepackage{courier}  
\usepackage[hyphens]{url}  
\usepackage{graphicx} 
\usepackage{natbib}  
\usepackage{caption} 
\usepackage{graphicx} 
\usepackage{times}  
\usepackage{helvet}  
\usepackage{courier}  
\usepackage[hyphens]{url}  
\usepackage{graphicx} 
\usepackage{natbib}  
\usepackage{caption} 
\usepackage[table,xcdraw]{xcolor}
\usepackage{hhline}
\usepackage{amsthm,amsmath,amssymb}
\usepackage{booktabs}
\usepackage{multirow} 
\DeclareCaptionStyle{ruled}
{labelfont=normalfont,labelsep=colon,strut=off} 
\usepackage{algorithmic}

\usepackage[linesnumbered,ruled,vlined]{algorithm2e}

\usepackage{newfloat}
\usepackage{listings}

\title{Prompting Vision Language Model with Knowledge from Large Language Model for Knowledge-Based VQA}
\author {
        Yang Zhou\textsuperscript{\rm 1,\rm 2}, Pengfei Cao\textsuperscript{\rm 1,\rm 2}, Yubo Chen\textsuperscript{\rm 1,\rm 2}, Kang Liu\textsuperscript{\rm 1,\rm 2}, Jun Zhao\textsuperscript{\rm 1,\rm 2}
}
\affiliations{
    \textsuperscript{\rm 1} The Laboratory of Cognition and Decision Intelligence for Complex Systems, Institute of Automation,\\ Chinese Academy of Sciences, Beijing, 100190, China\\
    \textsuperscript{\rm 2} School of Artificial Intelligence, University of Chinese Academy of Sciences,\\
    Beijing, 100049, China\\
    \{yang.zhou2020, pengfei.cao, yubo.chen, kliu, jzhao\}@nlpr.ia.ac.cn
}

\usepackage{bibentry}

\begin{document}

\maketitle
\begin{abstract}
Knowledge-based visual question answering is a very challenging and widely concerned task.
Previous methods adopt the implicit knowledge in large language models (LLM) to achieve excellent results, but we argue that existing methods may suffer from biasing understanding of the image and insufficient knowledge to solve the problem. 
In this paper, we propose PROOFREAD -\textbf{PRO}mpting vision language model with kn\textbf{O}wledge \textbf{F}rom la\textbf{R}g\textbf{E} l\textbf{A}nguage mo\textbf{D}el, a novel, lightweight and efficient knowledge-based VQA framework, which make the vision language model and the large language model cooperate to give full play to their respective strengths and bootstrap each other.
In detail, our proposed method uses LLM to obtain knowledge explicitly, uses the vision language model which can see the image to get the knowledge answer, and introduces knowledge perceiver to filter out knowledge that is harmful for getting the correct final answer.  
Experimental results on two datasets prove the effectiveness of our approach. 
Our method outperforms all state-of-the-art methods on the A-OKVQA dataset in two settings and also achieves relatively good performance on the OKVQA dataset.
\end{abstract}
\section{Introduction}

\begin{figure}[tbp]
    \centering
    \includegraphics[width=1\linewidth,scale=0.80]{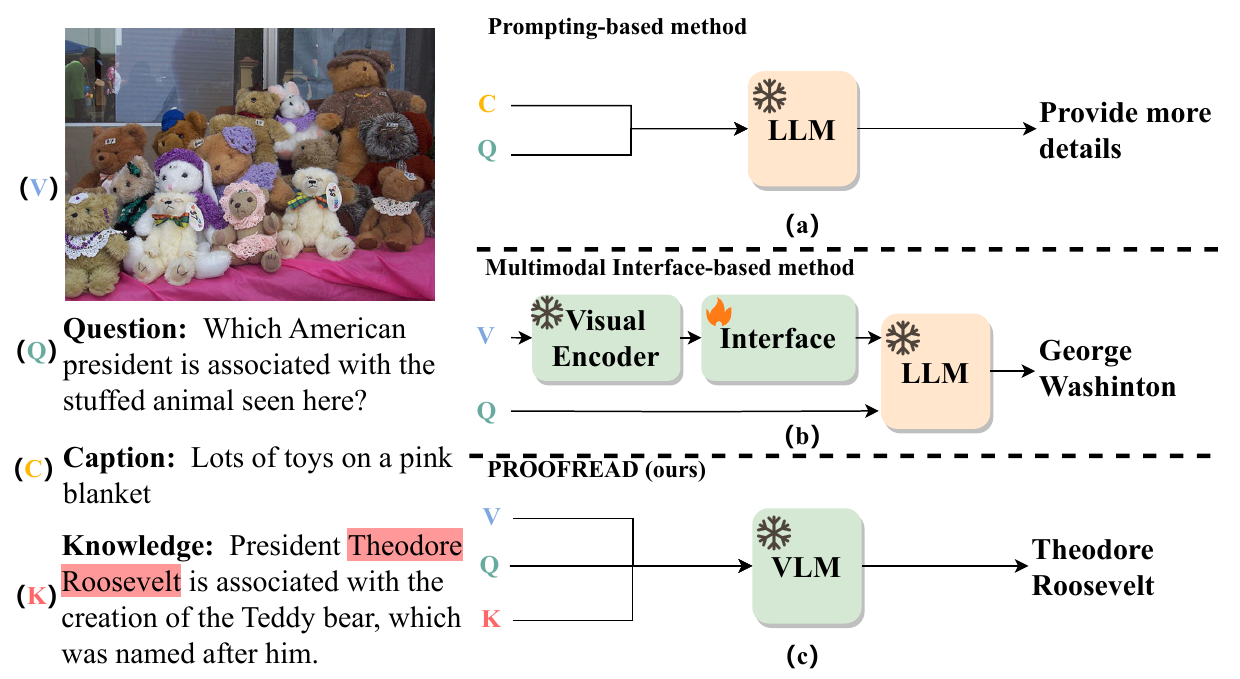}
     \caption{Examples of different approaches based on LLMs. Snowflakes represent that the module is frozen, and flames represent that the module is trainable. (a) An example of prompting-based methods, which convert images to captions fed into the LLM. (b) An example of Multimodal interface-based methods, which train a multimodal interface for large models that can understand image inputs. (c) Part of PROOFREAD (our method), the frozen VLM that leverages knowledge can answer questions correctly.  }
    \label{f1}
\end{figure}

Visual question answering (VQA) \cite{VQA/iccv/AntolALMBZP15} aims to allow machines to understand images and answer free-form questions by reasoning on given images. 
However, in real scenarios, it is not enough to answer questions only based on the content of images.
Machines also need to know some world knowledge or commonsense knowledge of the objects involved in the image.
Motivated by this, knowledge-based VQA \cite{fvqa/factvqa/oldbench,oldbench2,OKVQA,AOKVQA} has received increasing interest these years, which focuses on questions that require knowledge to answer.
For example, as shown in Figure~\ref{f1}, to answer the question ``\emph{Which American president is associated with the stuffed animal seen here?}'', it does not only need to recognize that the image is about teddy bears but also should know the relationship between ``\emph{teddy bear}'' and ``\emph{George Roosevelt}''  (George Roosevelt named the teddy bear).

Traditional methods \cite{old1/mukea/cvpr/DingYLHC022,old2/krisp/cvpr/MarinoCP0R21,old3//mm/YuCYWTT20,old4/validateKBVQA/aaai/WuLSM22,old5/Mucko/ijcai/ZhuYWS0W20} model this problem as a two-stage process, where the first stage retrieves knowledge from an explicit external knowledge resources (e.g., Wikipedia, ConceptNet \cite{conceptnet/aaai/SpeerCH17}), and the second stage fuses knowledge with images and questions to predict answers.
However, due to the incompletion of knowledge bases and the poor generalization of these methods which are limited by the training data scale, it is difficult for them to achieve excellent performance.
With the development of large language models (LLMs) \cite{GPT3/nips/BrownMRSKDNSSAA20,ChatGPT/nips/Ouyang0JAWMZASR22,chinchilla/deepmind}, in the field of natural language processing (NLP), some knowledge-intensive tasks (e.g., commonsense question answering \cite{commonsenseQA/naacl/TalmorHLB19}, open domain question answering \cite{ODQA/lrec/VoorheesT00}, document summarization \cite{documentsum/acl/SeeLM17} and etc.) no longer rely on external knowledge bases and can still achieve human-like performance.
Motivated by this, some studies \cite{PICA/aaai/YangGW0L0W22,KAT/naacl/GuiWH0BG22,Prophet,promptcap/corr/abs-2211-09699} attempt to take advantage of large amount of knowledge stored in the LLMs and its powerful reasoning ability to solve knowledge-based VQA.
However, since existing LLMs are basically trained on text, the information in the images cannot be directly understood by text-only LLMs (e.g., GPT-3 \cite{GPT3/nips/BrownMRSKDNSSAA20}, ChatGPT \cite{ChatGPT/nips/Ouyang0JAWMZASR22}). 
In order to make up for this gap, some methods have been proposed, which can be roughly divided into two categories: prompting-based methods, and multimodal interface-based methods.

\textbf{Prompting-based methods}: Some methods \cite{PICA/aaai/YangGW0L0W22,Prophet,promptcap/corr/abs-2211-09699} try to provide information about the image in the prompt to the LLM and predict the answer. 
For example, as illustrated in Figure \ref{f1}(a), PICa \cite{PICA/aaai/YangGW0L0W22} prompts a frozen GPT-3 with the caption of the image ``\emph{Lots of toys on a pink blanket}'' to provide the information of the image.
However, these methods may suffer from \textbf{biasing understanding of the image}, because the caption information of the image is not enough to elicit all the knowledge needed to solve the problem in the LLM.
In other words, the limitation of these methods is that the LLM cannot really see the image, and information (captions or answer candidates) is always difficult to exhaust all the key content in the image.
Under this condition, it is probably wrong to rely on the caption to prompt LLM to get the final answer.
For instance, the caption about the image in Figure~\ref{f1} is ``\emph{Lots of toys on a pink blanket}'', the contribution of which is very limited to help answer the question ``\emph{Which American president is associated with the stuffed animal seen here?}''.
With the caption as the prompt, it is difficult for LLMs to have an accurate and perfect understanding of image content, and then LLMs will be confused to obtain the answer. 

\textbf{Multimodal interface-based methods}: Some studies \cite{blip2/corr/abs-2301-12597,kosmos-2/corr/abs-2306-14824,flamingo/nips/AlayracDLMBHLMM22} attempt to train a multimodal interface for the LLM so that they can directly see the image to obtain the answer, which is also a very important way in constructing vision language model (VLM).
As Figure illustrated in \ref{f1}(b), this kind of VLM often freezes the image encoder and language model, and maps the image representation to the embedding space of the language model by training a multimodal interface on multimodal data.
In this framework, almost all knowledge is stored in the language model, if the language model does not contain the knowledge needed to solve the problem, then the question remains unanswerable.
In fact, existing VLMs that be publicly available to access still rely on relatively small-scale language models (e.g., FlanT5 \cite{FlanT5/corr/abs-2210-11416}, OPT \cite{OPT/corr/abs-2205-01068}), and the amount of knowledge stored in the model is limited.
In summary, these methods may suffer from \textbf{insufficient knowledge} to solve the knowledge-based VQA problem.
As shown in Figure~\ref{f1}(b), asking questions to the VLM directly will be answered incorrectly.
But as shown in Figure~\ref{f1}(c), with the help of appropriate knowledge, the problem can be solved.
And according to our sampling statistics, 43\% of the error of BLIP2 \cite{blip2/corr/abs-2301-12597} on the AOKVQA dataset can be corrected by providing appropriate knowledge.
This phenomenon shows that only relying on the implicit knowledge in the current VLM is not enough to answer the question.


Intuitively, in this paper, we propose \textbf{PROOFREAD} -\textbf{PRO}mpting vision language model with kn\textbf{O}wledge \textbf{F}rom la\textbf{R}g\textbf{E} l\textbf{A}nguage mo\textbf{D}el, a lightweight and efficient knowledge-based VQA framework.
Firstly, to avoid the \textbf{biasing understanding of the image}, PROOFREAD adopts a frozen VLM to predict the answers, which can see the image.
Secondly, to alleviate the \textbf{insufficient knowledge} of existing VLMs, PROOFREAD leverages an LLM to make up for this gap. 
Apart from making up for the weakness of the above methods, PROOFREAD also has the following two advantages. 
Firstly, PROOFREAD bridges LLMs and VLMs without a complex training process (Only 3000 samples were selected to adjust a small number of parameters), during which the parameters of LLM and VLM are frozen.
Secondly, PROOFREAD provides a paradigm for efficiently generating knowledge by large models, and proposes Knowledge Perceiver, a novel knowledge filtering mechanism, to minimize the impact of harmful knowledge.
Experimental results demonstrate the effectiveness of our proposed method.

Our contribution can be summarized as follows:
\begin{itemize}
\item We design a novel lightweight framework PROOFREAD to bridge LLM with the vision language model, which can help VLM with the knowledge from LLM, without complex training process.
\item  PROOFREAD can efficiently generate knowledge by LLMs.
The proposed knowledge perceiver can effectively filter harmful knowledge.
\item  We conduct extensive experiments on two public open-domain knowledge-based VQA datasets.
Experimental results prove the effectiveness of our method. 
Our code will be publicly available soon.
\end{itemize}

\begin{figure*}[t]
    \centering
    \includegraphics[width=0.80\linewidth,height=0.40\linewidth,scale=1]{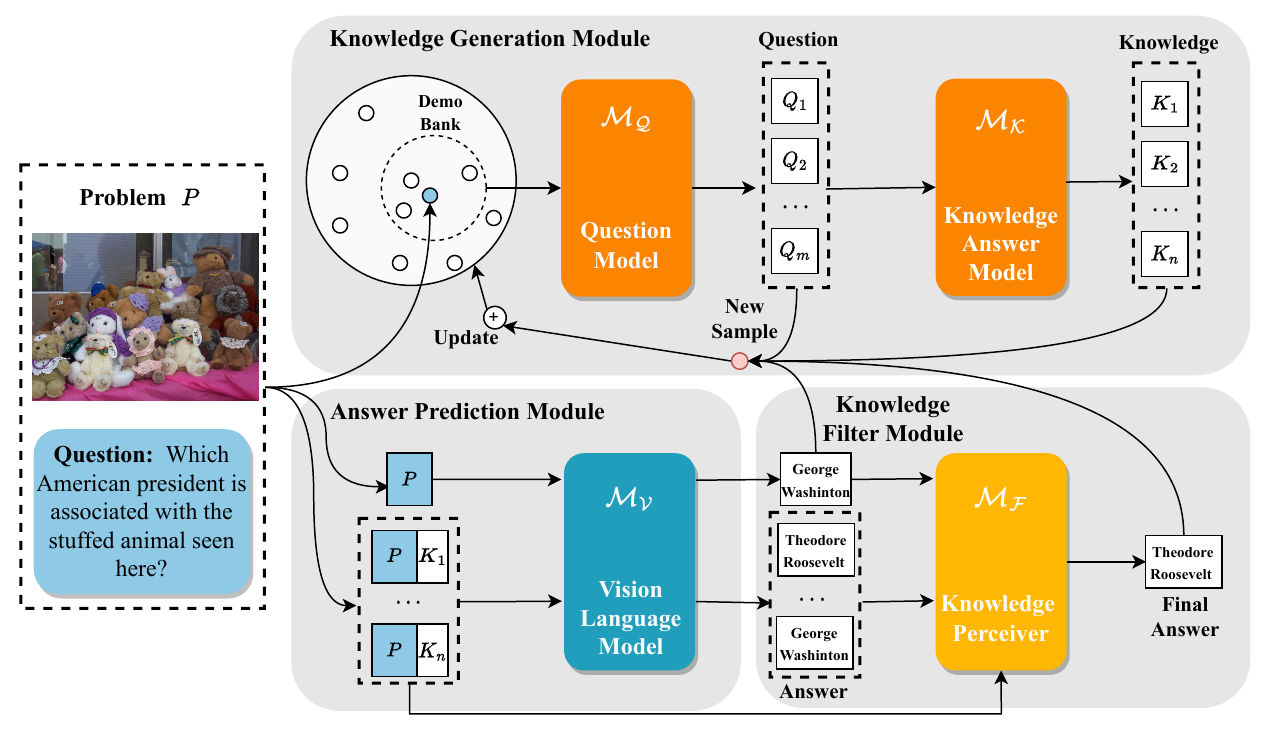}
\caption{The overall framework of \textbf{PROOFREAD}. Top: Knowledge Generation Module. Retrieve k similar examples from the Demo Bank according to the current problem ($P$), and generate m questions through the in-context learning of the Question Model ($\mathcal{M}_Q$). Generate n pieces of knowledge through the Knowledge Answer Model ($\mathcal{M}_K$ ). Bottom Left: Answer Prediction Module. Answer the questions about the image through the Vision Language Model ($\mathcal{M}_V$). Bottom Right: Knowledge Filter Module. According to the relationship between knowledge and questions, the final answer (e.g., Theodore Roosevelt) is given by Knowledge Perceiver ($\mathcal{M}_f$).}  
    \label{f_2_1}
\end{figure*}
\section{Related work}
\subsubsection{Knowledge-based Visual Question Answering.}
Knowledge-based visual question answering \cite{OKVQA,AOKVQA,fvqa/factvqa/oldbench,oldbench2} aims to solve questions that cannot be answered based on images alone and requires some knowledge or commonsense.
Early works rely heavily on retrieving knowledge from knowledge bases and then training to get answers.
Mucko \cite{old5/Mucko/ijcai/ZhuYWS0W20} first proposes to depict an image by a multi-modal heterogeneous graph containing multiple layers of information
based on visual, semantic, and knowledge modalities, and then answer questions.
KRISP \cite{old2/krisp/cvpr/MarinoCP0R21} combines implicit knowledge in Transformer and symbolic knowledge in the knowledge bases for the knowledge-based VQA.
However, due to the incompleteness of the existing knowledge resources, these methods are not able to achieve decent performance in the open domain knowledge-based VQA \cite{OKVQA,AOKVQA}.
Benefiting from the powerful reasoning ability and abundant knowledge of large language models(e.g., GPT-3 \cite{GPT3/nips/BrownMRSKDNSSAA20}, ChatGPT \cite{ChatGPT/nips/Ouyang0JAWMZASR22}), some studies \cite{flamingo/nips/AlayracDLMBHLMM22,PICA/aaai/YangGW0L0W22,Prophet,KAT/naacl/GuiWH0BG22,promptcap/corr/abs-2211-09699} utilize large language models as part of their methods and achieve excellent performance.
PICA \cite{PICA/aaai/YangGW0L0W22} converts images in knowledge-based VQA into captions and uses GPT3 and in-context learning to get the final answer.
Flamingo \cite{flamingo/nips/AlayracDLMBHLMM22} proposes a vision language model, which freezes the language model to save all language model capabilities, and trains the multimodal interface on the basis of it so that the model can cope with image input. 
Prophet \cite{Prophet} uses the VQA model to get candidate answers and then leverages GPT-3 to give the final answer through in-context learning.
Our method also exploits the large language model.

\subsubsection{In-Context Learning.}
The large language model, especially the large language model after instruction tuning, shows a powerful few-shot learning ability, giving some input and output demonstrations to the large language model, it can generate high-quality answers without training.
This training-free few-shot learning capability is called in-context learning.
And this capability also exists in multimodal large language models trained on multimodal data \cite{flamingo/nips/AlayracDLMBHLMM22,kosmos-2/corr/abs-2306-14824,minigpt4/corr/abs-2304-10592}.
Some studies\cite{icl1/make-good-example/acl-deelio/LiuSZDCC22,icl2/Reina/acl/WangXFLSX0022,icl3/learning-to-retrive/naacl/RubinHB22,ICL4/activelearning/emnlp/ZhangFT22,ICL5/unified/acl/LiLYLZNXWQ23} have shown that constructing a suitable example set can help enhance the ability of in-context learning.
REINA \cite{icl2/Reina/acl/WangXFLSX0022} retrieves samples similar to the test samples on the training set and then performs in-context learning to achieve good performance.
\citeauthor{ICL4/activelearning/emnlp/ZhangFT22} (\citeyear{ICL4/activelearning/emnlp/ZhangFT22}) propose a  reinforcement learning algorithm for identifying generalizable policies to select demonstration examples.
UDR \cite{ICL5/unified/acl/LiLYLZNXWQ23} proposes a unified model to retrieve demonstrations for a wide range of NLP tasks.

\renewcommand{\algorithmicrequire}{\textbf{Input:}}  
\renewcommand{\algorithmicensure}{\textbf{Output:}} 

\section{Methodology}

In this section, we will introduce the details of \textbf{PROOFREAD} framework for Knowledge-based VQA.
Figure~\ref{f_2_1} schematically visualizes our approach, which includes three major components: (1) Answer Prediction Module, which consists of a vision language model (VLM) for answering questions about the images based on the visual, question, and knowledge; (2) Knowledge Generation Module, which generates knowledge that may be needed to answer questions; (3) Knowledge Filter Module, which aims for filtering knowledge and giving the final answer.
Consequently, these three modules will be introduced in turn.

\subsection{Answer Prediction Module}
Answer Prediction Module adopts a frozen Vision Language Model($\mathcal{M_V}$, VLM) as the base model for Knowledge-based VQA because VLM has a certain ability to understand instructions as well as excellent performance.
Formally, define a VQA problem as $P = (v, q, a)$, where $v,q,a$ denotes the visual input, question about the image, and the answer to the question.
The original answer for VQA can be obtained by VLM
\begin{align}
    \hat{a}^o = \mathcal{M_V}(v,q) \in \mathbb{R}^V,
\end{align}
where $\hat{a}^o=\{\hat{a}^o_1, \hat{a}^o_2,...,\hat{a}^o_t\}$ is the original answer consisting of $t$ words and $\mathbb{R}^V$ is the answer space.
And the confidence score about the original answer of $\mathcal{M_V}$ is computed as follows:
\begin{equation}
p^o = \prod \limits_{i=1}^t p(\hat{a}^o_i |\hat{a}^o_1,..., \hat{a}^o_{i-1}) \in \mathbb{R},
\label{e2}
\end{equation}
where $p(\hat{a}^o_i |\hat{a}^o_1,..., \hat{a}^o_{i-1})$ denotes the prediction probability of the word $\hat{a}^o_i$.

Relying on image information alone cannot solve knowledge-based VQA, and some knowledge needs to be acquired at this time.
We design a novel Knowledge Generation Module to obtain knowledge, which will be introduced in the next section.
Knowledge Generation Module generates a series piece of knowledge $K = \{ K_1,K_2,...,K_n\}$ about the problem $P$.
Therefore, our framework combines every piece of knowledge and the question into prompts through manually designed templates and inputs them into VLM to obtain knowledge-based answers,
\begin{equation}
    \hat{a}^{ki} = \mathcal{M_V}(v,E_1(K_i,q)) \in \mathbb{R}^V,
\end{equation}
where $\hat{a}^ki=\{\hat{a}^{ki}_1, \hat{a}^{ki}_2,...,\hat{a}^{ki}_t\}$ is the knowledge answer consisting of $t$ words, and $E_1(\cdot,\cdot)$ is the prompt function to combine question and knowledge.
Similarly, the confidence score $p^{ki}$ about the knowledge answer $a^{ki}$ is obtained via Equation \ref{e2}.

\subsection{Knowledge Generation Module}

The goal of the Knowledge Generation Module is to generate knowledge for VQA efficiently and correctly.
Our framework leverages ChatGPT \cite{ChatGPT/nips/Ouyang0JAWMZASR22}, a language model with a large amount of parameters and strong interaction ability, to provide knowledge.
However, directly informing ChatGPT questions and letting it generate relevant knowledge will lead to subjective inferences in the generated knowledge.
The model often gives the guessed answer to the question first and gives the knowledge related to the answer it thinks, which results in limited knowledge generation.
Besides, LLMs may suffer from the biasing understanding of the image, and the answer it gives will also be unreliable, which leads to misleading knowledge generation.
Based on this, in order to generate knowledge more effectively, we set a Question Model ($\mathcal{M_Q}$) to generate relevant knowledge questions about VQA Problem $P$, and then use a Knowledge Answer Model ($\mathcal{M_K}$) to answer these questions.
We decouple the generation process into a question-and-answer format.
When generating knowledge questions about the problems, the subjective intention of the model in the generation process will be reduced.
It is very important to build high-quality demonstrations in in-context learning \cite{GPT3/nips/BrownMRSKDNSSAA20} for generating knowledge questions.
Our method proposes a new way to select high-quality examples.
First, we construct a Demo Bank $D = \{ d_i \}^{\tau}_{i=1}$ from the training set for in-context learning, which contains $\tau$ samples.
Then for new samples during training or testing,  demonstrations are selected from the Demo Bank to acquire knowledge.

PROOFREAD builds the Demo Bank on the training set.
We first randomly select a few samples from the training set.
Then ask some knowledge questions manually, and get the corresponding knowledge from Knowledge Answer Model $\mathcal{M_K}$.
And we adopt the Vision Language Model $\mathcal{M_V}$ to get the original answer and the knowledge answer for these samples, using which as the initialization seed of the Demo Bank.

For a testing or training sample $P = (v,q,a)$, our framework leverages the encoder of Vision Language Model ($\mathcal{M_V}$) to obtain the representation of the sample
\begin{equation}
    \mathbf{r} = Encoder_\mathcal{M_V}(v,q) \in \mathbb{R}^d,
\end{equation}
where $d$ stands for hidden size.
According to the representation of the sample, we will find the $k$ most similar samples from the Demo Bank $D$ as the demonstrations $D_p$ for generating the knowledge question, which can be formulated as 
\begin{align}
    S_{QE} &= \mathop{argTopK}\limits_{i \in \{1,2,...,\tau \} } \frac{\mathbf{r}^T  \mathbf{r}_i}{||\mathbf{r}||_2 ||\mathbf{r}_i||_2}, \\ 
    D_p &= \{(v_i,q_i,a_i)| i \in S_{QE}\}.
\end{align}
Constructing these demonstrations and the current problem into a prompt, we can obtain the knowledge questions generated by Question Model $\mathcal{M_Q}$ as
\begin{align}
    Q &= \{ Q_i \}_{i = 1} ^m \notag \\ 
    &= \mathcal{M_Q} (E_2(\{(C(v_i),q_i)| i \in S_{QE}\},(C(v),q))),
\end{align}
where $m$ is the number of generated questions, $E_2(\cdot,\cdot)$ is the prompt function to combine demonstrations and the current problem, $C(\cdot)$ denotes the caption function to convert the image to caption.
Finally, we feed the knowledge questions into Knowledge Answer Model $\mathcal{M_K}$ to get the knowledge for the problem $P$
\begin{align}
    K = \{ K_i \}_{i = 1} ^n = \mathcal{M_K} (E_3(Q)),
\end{align}
where $n$ is the number of knowledge, $E_3(\cdot)$ is the prompt function for ask knowledge questions to $\mathcal{M_K}$.
It is worth noting that $n$ is not equal to $m$, because in the prompt, we tell the model answer in points, and each question may generate multiple pieces of knowledge.

In order to evaluate the quality of automatically generated questions and knowledge on the training set, we categorize the generated knowledge into three classes: 
\begin{itemize}
    \item Useful Knowledge, which makes up for the lack of knowledge that the model has to solve the problem.
    \item Harmful Knowledge, which can mislead the model into giving wrong answers to otherwise solvable problems
    \item Neutural Knowledge, which has no appreciable impact on the model's ability to solve the problem.
\end{itemize}
Therefore, for a sample, the number of useful knowledge generated can be recorded as $u$, and the number of harmful knowledge is $h$
\begin{align}
    u &=  \sum_{i = 1}^ n \mathbb{I}(\hat{a}^{ki} = a)  \mathbb{I}(\hat{a}^{o} \ne a),  \\ 
    h &= \sum_{i = 1}^ n \mathbb{I}(\hat{a}^{ki} \ne a)  \mathbb{I}(\hat{a}^{o} = a),
\end{align}
where $\mathbb{I}(\cdot)$ denotes indicator function, $n$ is the number of knowledge, $a^{ki}$ is the i-th knowledge answer and $a^o$ is the original answer, $a$ is the ground truth answer. 
Traverse each sample on the training set, find similar samples from the Demo Bank to generate knowledge, get the answers to the problem, and calculate the $u$ and $h$ values.
For each newly generated sample, the process to update the DEMO Bank is shown in the Algorithm \ref{alg1}.
In order to ensure the diversity of samples in the Demo Bank and to retrieve them efficiently, each time a sample whose similarity with the generated sample is greater than a certain threshold $\lambda$ is found from the Bank, the two are compared, and the excellent samples are retained.


\begin{algorithm}[t] 
\small
\caption{Demo Bank Update Algorithm} 
\label{alg1} 
\begin{algorithmic}[1] 
\REQUIRE Current Demo Bank $D = \{d_i\}_{j=1}^n =\{(\mathbf{r}_i, v_i, q_i, a_i, \{K_i^j\}_{j=1}^n,\hat{a}^o_i, \{\hat{a}^{kj}_i\}_{j=1}^n),h_i,u_i\}_{i=1}^t$, similarity threshold $\lambda$, new sample $s = (\mathbf{r}, v, q, a, \{K^j\}_{j=1}^n,\hat{a}^o, \{\hat{a}^{kj}\}_{j=1}^n,h,u)$  
\FOR{$d_i$ in $D$}

\IF{$\frac{\mathbf{r}^T  \mathbf{r}_i}{||\mathbf{r}||_2 ||\mathbf{r}_i||_2} > \lambda$} 
\IF{$h < h_i$} 

\STATE remove $d_i$ from $D$ 
\STATE add $s$ to $D$
\ELSIF{ $h = h_i$ and $u > u_i$}
\STATE remove $d_i$ from $D$
\STATE add $s$ to $D$
\ENDIF 
\ENDIF 
\ENDFOR
\ENSURE New Demo Bank $D$ 
\end{algorithmic} 
\end{algorithm}

\subsection{Knowledge Filter Module}
This module aims to filter out useful knowledge from the generated knowledge, which can also be defined as classifying knowledge into three classes (useful, harmful, and neutral), and give the final answer according to the classification of knowledge.
Our method leverages XGBoost \cite{xgboost/kdd/ChenG16}, a gradient-boosted decision tree (GBDT), as the Knowledge Perceiver $\mathcal{M_F}$ to classify the knowledge. 
For a knowledge-based VQA problem $P = (v,q,a)$
We selected 11 relevant features to classify the knowledge, which is introduced as follows:
\begin{itemize}
\item \textbf{Original Confident}. The confident score $p^o$ about the original answer $a^o$ from $\mathcal{M_V}$.
\item \textbf{Knowledge Confident}. The confidence score $p^{ki}$ about the knowledge answer $a^{ki}$ from $\mathcal{M_V}$.
\item \textbf{Confident Gain}. The difference between Knowledge Confident and Original Confident, which can be 
 formally defined as $p^{ki} - p^{o}$.
\item \textbf{Text Similarity}. The cosine similarity $s^{ti}$ of the representation between the question about the image and a piece of knowledge. 
\item \textbf{Caption Similarity}. Cosine similarity $s^{ci}$ between the image caption representation and the knowledge representation.
\item \textbf{Image Similarity}. Cosine similarity $s^{vi}$ between the image representation and the knowledge representation. 
\item \textbf{Entailment}. The entailment score $e$ between the question and knowledge. 
can be obtained by $ ent = Ent(q,K_i)$, where $Ent(\cdot,\cdot)$ is the entailment model.
\item \textbf{Contradiction}. The contradiction between the question and knowledge. 
can be obtained similarly by $ con = Ent(q,K_i)$ 
\item \textbf{Knowledge Confident Important}. Among all the knowledge of the problem, $K = \{ K_1,K_2,...,K_n\}$, 
the importance $I^C$ of a certain piece of knowledge $K_i$ is the softmax of $p^{ki}$.
from the perspective of knowledge confidence.
\item \textbf{Knowledge Visual Important}. Similarly, among all the knowledge of the problem, the importance $I^v$ of a certain piece of knowledge is the softmax of $s^{vi}$,
from the perspective of the similarity between knowledge and image.
\item \textbf{Knowledge Caption Important}. Among all the knowledge of the problem, the importance $I^{kc}$ of a certain piece of knowledge is the softmax of $s^{ci}$,
from the perspective of the similarity between knowledge and image caption.
\end{itemize}

For each piece of knowledge during the test process, the knowledge will be classified by the knowledge perceiver, and finally, the answer with the highest vote of useful knowledge and netural knowledge will be adopted as the final answer.

\section{Experiment}

\subsection{Experimental Settings}
\subsubsection{Datasets.}
We choose two public knowledge-based VQA datasets \textbf{OKVQA} \cite{OKVQA} and \textbf{A-OKVQA} \cite{AOKVQA} to validate the effectiveness of our method.
\textbf{OKVQA} is a large knowledge-based VQA dataset, which contains 14k question-answer pairs and 14k images from the MSCOCO dataset \cite{MSCOCO/eccv/LinMBHPRDZ14}.
Questions in the dataset written by annotators during construction are required to contain knowledge.
\textbf{A-OKVQA} is currently the largest knowledge-based VQA benchmark, which is an augmented successor of OK-VQA and contains a diverse set of 25k questions requiring a broad base of knowledge to answer. 
For the two datasets, we randomly sample 3k samples from their training sets as training samples for building Demo Bank and training Knowledge Perceiver.

\subsubsection{Implementation Details.}
We adopt frozen BLIP-2 (FlanT5XXL) \cite{blip2/corr/abs-2301-12597} as Vision Language Model 
to generate the answer and generate the caption of the image.
For multiple-choice output, we constrain the model to only generate option words (i.e., \emph{abcd}).
The demonstration number $k$ used when generating knowledge questions is set to 3.
When constructing the Demo Bank, our framework traverses the training set twice.
The proposed method leverages MPNET \cite{MPNET/nips/Song0QLL20} for representing the text to compute the similarity between questions and knowledge or captions and knowledge.
The entailment and contradiction score adopts the entailment model used by \citeauthor{Entailment/emnlp/HonovichCANSA21} (\citeyear{Entailment/emnlp/HonovichCANSA21}).
The similarity between the image and text is calculated using EVA-CLIP\cite{eva-clip/corr/abs-2211-07636}.
All experiments are conducted with NVIDIA GeForce RTX 3090 GPUs.

\begin{table}[t]
\centering
\small  
\begin{tabular}{lcccc}
\toprule
\multirow{2}{*}{\textbf{Method}} & \multicolumn{2}{c}{\textbf{Direct Answer}} & \multicolumn{2}{c}{\textbf{Multiple Choice}} \\
                        & \textbf{val}             & \textbf{test}            & \textbf{val}             & \textbf{test}            \\ \midrule
\specialrule{0em}{1pt}{1pt}
\midrule
\multicolumn{5}{c}{ \emph{conventional methods}}              \\
Pythia                              & 25.2            & 21.9   & 49.0            & 40.1         \\
ViLBERT                             & 30.6            & 25.9   & 49.1            & 41.5         \\
LXMERT                              & 30.7            & 25.9   & 51.4            & 41.6         \\
KRISP                               & 33.7            & 27.1   & 51.9            & 42.2         \\
GPV-2                               & 48.6            & 40.7   & 60.3            & 53.7         \\
BLIP-2                              & 53.2            & 49.7   & 70.2            & 69.4         \\ \midrule
\multicolumn{5}{c}{ \emph{large language model-based approaches}}              \\
ClipCap                            & 18.1            & 15.8    & 44.0            & 43.8         \\
Phrophet                           & 58.2            & 55.7    & 76.4            & 73.6         \\
PromptCap                          & 56.3            & 59.6    & 73.2            & 73.1         \\ \midrule
\textbf{PROOFREAD}                           & \textbf{62.2}            & \textbf{60.2}  & \textbf{77.8}            & \textbf{76.1}       \\ \bottomrule
\end{tabular}
  \caption{Overall performance (accuracy) comparison in A-OKVQA dataset.}
  \label{t1}
\end{table}

\begin{table}[t]
  \centering
  \small 
\begin{tabular}{lr}
 
\toprule \\
\textbf{Method}                       & \textbf{Accuracy}                 \\ \midrule
\specialrule{0em}{1pt}{1pt}
\midrule
\multicolumn{2}{c}{ \emph{conventional methods}}              \\
Mucko                     & 29.2                        \\
ConceptBERT               & 33.7                        \\
KRISP                     & 38.9                        \\
MAVEx                     & 40.3                        \\
TRiG                      & 49.4                        \\ 
BLIP-2                   & 45.9                        \\ \hline
\multicolumn{2}{c}{\emph{~~~~~~~~~large language model-based approaches}  ~~~~~~~~~} \\
PICA                      & 48.0                        \\
KAT                       & 53.1                        \\
Flamingo-80B              & 57.8                        \\
Phrophet                  & 61.1                        \\ \midrule
\textbf{PROOFREAD}                 & 57.6    \\
\bottomrule
\end{tabular}
  \caption{Overall performance comparison in OKVQA dataset. The upper part is the methods based on the knowledge base, and the lower part is the methods leveraging a large language model.}
   \label{t2}
\end{table}

\subsection{Comparisons with SOTA methods}
As shown in Table ~\ref{t1}, our method outperforms other methods on the A-OKVQA data set, whether in the setting of generating the answer directly or giving options to choose the final answer.
Especially under the setting of multiple choice, our method can outperform the state-of-the-art (SOTA) methods by a large margin with 3.0\% accuracy in the test set.
In addition, from the Table \ref{t1}, we notice:

(1) Our method stands out among large language model-based methods, indicating that our method indeed combines the vision language model's capability of understanding images and the knowledge of advanced large models.

(2) And it is worth mentioning that our method is based on BLIP-2 \cite{blip2/corr/abs-2301-12597}, our method outperforms BLIP-2 with 6.7\% accuracy in the setting of multiple choice and 10.4\% accuracy in the setting of direct answer in the test set of A-OKVQA, which validates that our framework can exploit the knowledge.

(2) Large language model-based approaches are always performing better than conventional methods.  
This phenomenon illustrates the superiority of large language models in answering knowledge/commonsense questions.

Similarly, as shown in Table ~\ref{t2} our method also achieves relatively good performance on OKVQA.
Although not exceeding all baselines, our framework training is very efficient. 
Only training on a small number of samples, and freezing the parameters of the vision language model, our method can achieve a similar performance to the baseline after multi-modal pre-training or the baseline fully trained on the OKVQA dataset.
In addition, we guess that our method does not exceed some baselines because few parameters cannot fully fit the dataset.

\begin{table}[t]
  \centering
  \small  
\begin{tabular}{lc}
\toprule
\textbf{Method}      & \textbf{Accuracy} \\ \midrule
\specialrule{0em}{1pt}{1pt}
\midrule
\textbf{PROOFREAD}   & \textbf{77.8}              \\ \midrule
w/o Demo Bank           & 76.2              \\
w/o Knowledge Perceiver ~~~~~~~~~~~~~~~ & 71.4              \\
w/o Knowledge           & 70.2              \\ \bottomrule
\end{tabular}
\caption{Ablation study of \textbf{PROOFREAD} in A-OKVQA validation set.}
   \label{t3}
\end{table}

\begin{figure*}[t]
    \centering
    \includegraphics[width=0.99\linewidth, height=0.25\linewidth, scale=1]{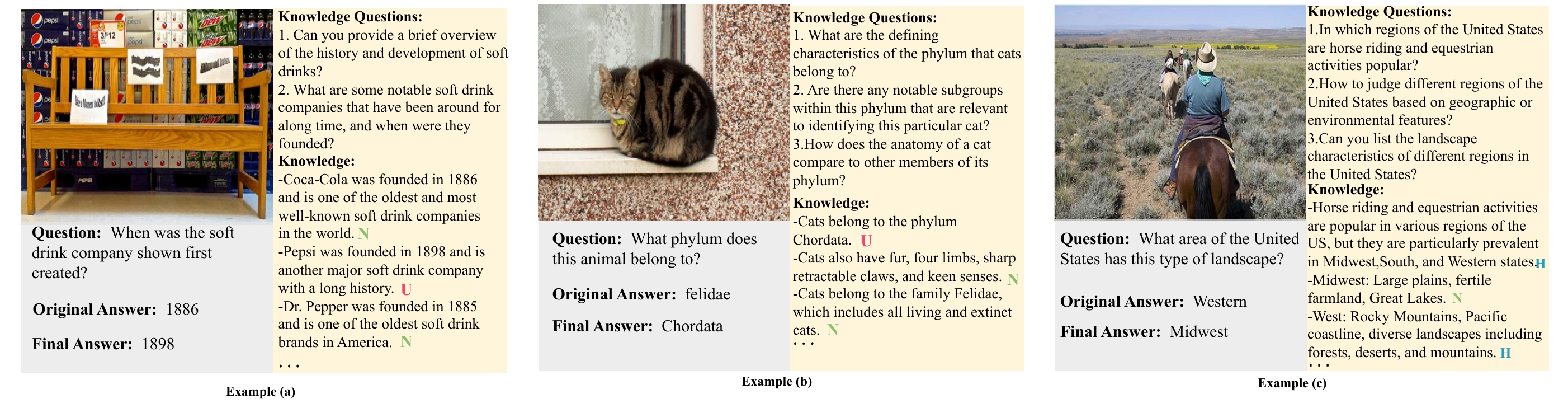}
\caption{Three cases of \textbf{PROOFREAD} on OKQVA dataset and A-OKVQA dataset. For each example, the left side shows the image, the question, the predicted answer without using knowledge, and the predicted final answer using knowledge, while the right side shows the knowledge questions and knowledge generated according to the questions, where \textcolor{magenta}{U}, \textcolor{green}{N}, \textcolor{teal}{H} represent the ground truth label for the knowledge being useful, neutral, or harmful respectively.}
    \label{f_3_2}
\end{figure*}

\begin{figure}[t]
    \centering
    \includegraphics[width=0.99\linewidth,scale=1]{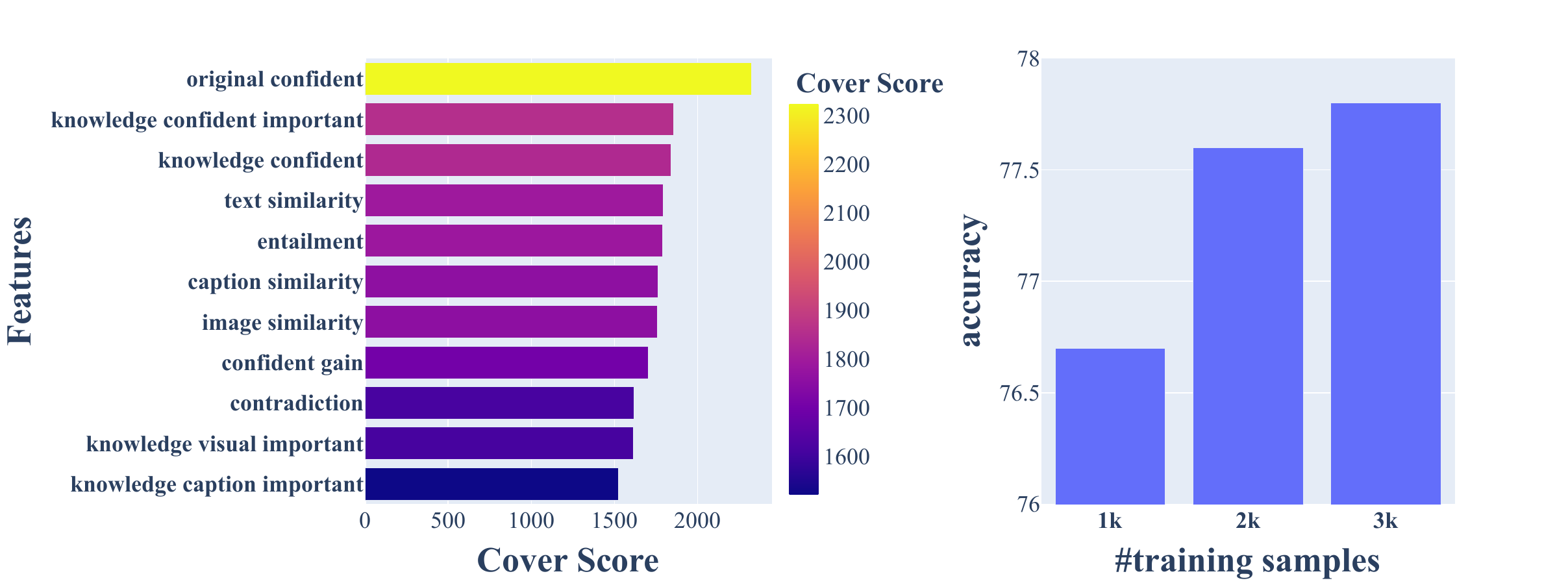}
\caption{\textbf{Left}: The importance of different features for knowledge perceiver in A-OKVQA validation set. The cover rate represents the number of samples that different features can cover in a gradient-boosted decision tree (GBDT), which represents the importance of this feature for knowledge classification.
\textbf{Right}: The effect of using different numbers of training samples for the final result in the A-OKVQA val set.
}
    \label{f_3_1}
\end{figure}

\subsection{Ablation Study}
To verify the effectiveness of each component of PROOFREAD, we conduct ablation studies on A-OKVQA.  Here we consider three settings: 
\begin{itemize}
\item {w/o Demo Bank}, which use fixed manual demonstrations to generate knowledge questions.
\item {w/o Knowledge Perveiver}, which without filtering, use the answer that the model gives the highest confidence, as the final answer.
\item {w/o Knowledge}, which does not use any external knowledge (i.e., BLIP-2).
\end{itemize}
It can be seen from Table \ref{t3} that each module is effective, and then we will analyze them one by one.

\subsubsection{Effectiveness of Demo Bank.} 
When adopting fixed manual samples as demonstrations for generating knowledge questions, we can see that the performance has dropped by 1.6\% accuracy.
It shows that PROOFREAD selects samples from the updateable Demo Bank for in-context learning does improve the quality of generating knowledge questions.

\subsubsection{Effectiveness of Knowledge Perceiver.} 
When replacing the knowledge perceiver with the direct usage of the knowledge answer with the model's highest confidence as the final answer, performance drops dramatically (6.4\% accuracy).
Our analysis may be because there will be some misleading knowledge during the generating process, which may make the model mistake the original correct answer and be very confident. 
This can also be regarded as a phenomenon of hallucination of the model.
Thus, it is very important to filter out harmful knowledge for knowledge-based VQA.

\subsubsection{Effectiveness of Knowledge.} 
Our framework is equivalent to the Vision Language Model (BLIP-2) if no knowledge is used at all.
At this time, the performance dropped by 7.6\%, which shows that the Vision Language Model is indeed lacking in knowledge. 
If the gap can be supplemented by a large language model, even without complex training, it can achieve good performance.

\subsection{Discussion and Analysis}

We also explore the importance of different features for filtering knowledge and the impact of different numbers of training samples on the final performance.

\subsubsection{Importance of different features.} 
As illustrated in Figure \ref{f_3_1} left, the importance of different features for classifying knowledge in the A-OKVQA validation set, the most important feature is the model confidence of for the original answer.
In classification, we find that if the vision language model has low confidence in the original answer, it is often because it lacks the knowledge to solve the problem.
Knowledge confidence important represents the importance of the model to obtain answers through different knowledge. 
The more important the knowledge, the higher the confidence that can often be obtained.

\subsubsection{Impact of different numbers of training samples.}
Our method leverages part of the training set to adjust the parameters, and the results obtained with different sample numbers in the A-OKVQA validation set are shown on the right side of Figure \ref{f_3_1}.
Using only 1k samples, 76.7\% accuracy can already be achieved, which illustrates the effectiveness of our framework for data utilization.
Intuitively, as the amount of data used rises, the performance of our method also increases.

\subsection{Case Study}

To illustrate the intermediate results of our method more intuitively, we pick some examples to illustrate.
Figure \ref{f_3_2} shows three cases of our method on the OKVQA and A-OKVQA datasets, as well as the knowledge questions generated and the knowledge obtained (partially).
In Example (a) and (b), answering the question directly by the vision language model leads to a wrong answer, due to lacking of knowledge.
In Example (a), directly answering the creation time of Pepsi-Cola Company will be incorrectly answered as the creation time of Coca-Cola, probably because of the lack of knowledge of the creation time of Pepsi-Cola.
With the blessing of useful knowledge, the vision language model can answer correctly.
In Example (a), after getting the useful knowledge of the creation time of Pepsi, our method can answer it correctly.
In both Example (a) and Example (b) our method effectively utilizes useful knowledge to get the correct final answer.

However, our method also has limitations. 
The introduction of knowledge may also introduce harmful information and turn the originally correct answer into a wrong one.
For instance, in Example (c) our method generates some harmful knowledge and adopts the answer generated by the wrong knowledge as the final answer.
At this time, it is necessary to effectively filter the knowledge. 
The previous experiments have verified the effectiveness of our method in filtering knowledge.

\section{Conclusion}
In this paper, we propose a novel and efficient framework to combine the vision language model and the large language model for  knowledge-based VQA.
The proposed method uses a large language model to acquire knowledge and a vision language model to answer questions.
To generate high-quality knowledge, we propose to use an updatable Demo Bank to pose knowledge questions.
To reduce the impact of harmful knowledge, we propose to use Knowledge Perceiver to filter knowledge.
In future work, we will explore the possibility of extending PROOFREAD to more knowledge-intensive tasks.

\bibliography{aaai22}
\bibliographystyle{aaai24}

\end{document}